\documentstyle[12pt]{article} 
\begin{document}
%\twocolumn[\hsize\textwidth\columnwidth\hsize\csname@twocolumnfalse%
%\endcsname
\title{ Stochastic Model in the Kardar-Parisi-Zhang Universality With Minimal
Finite Size Effects }  
\author{S.V. Ghaisas \\  
%\address{
 Department of Electronic Science, University of Pune,\\
 Pune 411007,India
}

%\date{\today}
                  
\maketitle 

%\begin{abstract} 
  We introduce a solid on solid lattice model for growth with conditional 
evaporation. A measure of finite size effects is obtained by observing the 
time invariance of distribution of local height fluctuations. The model 
parameters are chosen so that the change in the distribution in time is minimum.
 On a one dimensional substrate the results obtained from the 
model for the roughness exponent $\alpha$ from three different methods are 
same as predicted for the Kardar-Parisi-Zhang (KPZ) equation. One of the 
unique feature of the model is that the $\alpha$ as obtained from the structure 
factor $S(k,t)$ for the one dimensional substrate growth exactly matches with 
the predicted value of 0.5 within statistical errors. The model can be defined 
in any dimensions. We have obtained results for this model on a 2 and 3 dimensional 
substrates. 

%\end{abstract} 

%\pacs{PACS numbers: 
{pacs}{60., 68.55-a,82.20.Fd}   
\vfill
\pagebreak                          
%] 
%\narrowtext
  The KPZ equation \cite{kpz} is one of the most studied stochastic equation in the 
field of growth. 
\begin{equation}
    \frac{\partial h}{\partial t}=\nu_{0}\nabla^{2}h+ \lambda(\nabla h)^{2}
                     +  \eta  
\end{equation}
Here, $h(r,t)$ is the height function, $\lambda$ is coupling parameter, and 
$\eta(r,t)$ is Gaussian noise with the correlation $<\eta(r,t)\eta(r',t')>
=\delta(r-r',t-t')$. First term is a linear term \cite{ew} referred as Edward-Wilkinson (EW)
term.  Applications and uses of KPZ equation have been well 
demonstrated \cite{st}. In particular its use in understanding  growth 
phenomenon has lead to vigorous activities in development of theoretical methods
\cite{ml,cast}, lattice models \cite{kk,wk,bd} and numerical methods \cite{ggr}
built around Eq. (1). 
The critical exponents in (1+1) dimensions are exactly obtainable \cite{bar}. However, 
in all the higher dimensions the determination of these exponents has been a 
difficult task. It is well known that this equation shows phase transition in 
dimensions higher than its critical dimension (2+1) \cite{bar} as a function 
of its coupling parameter. For weak coupling case the coupling constant renormalizes to zero
 leading to a linear equation. For strong coupling case, the perturbation approach 
fails and other methods are required. However, obtaining exact values for the 
exponents has never been possible although the ranges in which exact values are  
expected to fall are evident from the available references \cite{ml,cast,kk,wk,bd,ggr}. 

Various lattice models have been devised \cite{kk,wk,bd} that are known to belong 
to the KPZ universality in the asymptotic region. Most of the models suffer 
from finite size effects arising from cutoff length $a$ and the substrate size $L$. 
One of the tests to probe the presence of finite size effects is to determine the
growth exponents by different methods such as height-height (h-h) correlations, 
structure factor $S(k,t)=<h(k,t)h(-k,t)>$, from saturated widths $W_{sat}$, etc..
For any given lattice model, the values 
obtainable from these methods can be statistically different. Since all these models are 
expected to converge asymptotically to KPZ behavior, the apparent mismatch of the exponent 
values from different methods will be due to the finite size effects. On the other hand if a 
model gives same exponents within statistical errors, it is expected to be free of finite 
size effects. In the following we propose 
a model that we believe to belong to the KPZ universality and, in (1+1) dimensions, 
it provides the values of the exponent $\alpha$ same within the statistical error, 
using three different methods of determination. This value also compares well with the 
exactly known value in (1+1) dimensions, 0.5.    

We describe the model below and the changes therein for (2+1) and (3+1) dimensions.
A site is chosen randomly and height at the site is increased by unity signaling
random deposition on the substrate. The deposited atom is conditionally accomodated, otherwise   
evaporated. In (1+1) dimensions the deposited atom is accommodated 
if both its neighbors have at least same height as the deposited one. Otherwise, 
largest of the height differences at the site $i$ and the nearest neighboring sites 
,$s_{d}=max(h_{i}-h_{j}), j=i+1,i-1$, 
is obtained and accommodation is 
allowed according to the probability factor $e^{-s_{d}^2/(2\sigma ^{2})}$. Thus 
$s_{d}$ is the largest local step.  
Choice of $\sigma$ depends upon the behavior of the model for the given value of $\sigma$.
We choose the value of $\sigma$ that leads to minimum variation in the local height fluctuations. 
Model with such a $\sigma$ is expected to be least affected by the finite size effects \cite{sve1}. 
It has been shown in reference 
\cite{sve1} that a measure of finite size effects for a given lattice model can be obtained 
from the distribution of local height fluctuations. In this method, we define the local height 
$(h_{i}(t))_{local}=h_{i}(t)-(h_{i-1}(t)+h_{i+1}(t))/2$ with respect to the local reference as the 
average of nearest neighbor heights. Similarly we measure $(h_{i}(t+\Delta t))_{local}$ where 
$\Delta t> w(t)$. $w(t)$ is the width of the interface at $t$, and the inequality ensures 
that the difference between local heights measured at $t$ and $t+\Delta t$ are uncorrelated. 
Thus we measure the distribution of uncorrelated fluctuations $\Delta h(t)_{local}$ from the 
difference $\Delta h(t)_{local}=(h_{i}(t))_{local}-(h_{i}(t+\Delta t))_{local}$ . 
Fig. 1 shows such a distribution for $\sigma =1.7$. Another distribution is obtained 
at later time and compared with the earlier one. In our case we have obtained distributions 
at $t=$500 MLs and $t=$5000 MLs for comparison with $\Delta t=100$ MLs. 
Since the counts at $\Delta h(t)_{local}=0$ are 
largest in the distribution, the statistical error is minimum for zero fluctuation. We therefore 
use the parameter  $P_{0}=100(\frac{I_{500}-I_{5000}}
{I_{500}})$, where  $I_{t}$ is the count at $\Delta h(t)_{local}=0$ to measure the time 
invariance of the distribution of $\Delta h(t)_{local}$ in (1+1) and (2+1) dimensions. 
In (3+1) dimensions, $\Delta t$ and times for comparison are smaller due to the 
large computation times involved. Ideally $P_{0}$ should be zero. 
In the present context we look for a minimum value of $P_{0}$ as a function of $\sigma$.   
The ratios $P_{0}$ are obtained by averaging
over large enough runs so that values of $P_{0}$ are statistically discriminated for 
different values of $\sigma$. It has been shown in reference \cite{sve1} that this 
method is useful in identifying presence of finite size effects for any lattice model 
belonging to KPZ or EW universality. Fig. 2 shows the variation of $P_{0}$ with $\sigma$ 
for the model described. We have measured $P_{0}$ for the model with different $\sigma$ values 
on a substrate of length $L=40000$. As can be seen, the minimum occurs at  $\sigma=1.7$. We have 
therefore used this value in (1+1) dimensions. For other values of  $\sigma$ we found that 
value of $\alpha$ as obtained from structure factor deviates from 0.5 and the linear range 
is also reduced on the log-log plot. This confirms the effectiveness of the method for 
determining finite size effects \cite{sve1}.     
 
In (2+1) dimensions the deposited 
atom is accommodated if three or more nearest neighbors have at least same height as its own.
If this condition is relaxed to less number of in-plane neighbors, cross over due to 
EW region is obtained.The cross over is negligible when direct accommodation with
 three or four in-plane neighbors is allowed.   
For depositions at the site with less than three in-plane neighbors, 
the accommodation is decided from the largest of the four steps 
around the site using the exponential probability factor $e^{-s_{d}^2/(2\sigma ^{2})}$.
 In (2+1) dimensions we have 
observed that $\sigma=2.5$ shows minimum $P_{0}=0.002\pm 0.0015 \%$. The substrate size 
is $L=400$.  

In (3+1) dimensions the deposited atom is accommodated if 5 or 6 nearest neighbors have 
at least same height as the deposited atom. We have used $\sigma=4.5$ in the simulations
since this value gives minimum $P_{0}=0.0012\pm 0.001\%$. The $\Delta t=20MLs$ and the 
distributions are compared for the times 50 MLs and 500 MLs for the substrate size of 
$L=100$.   

 We present results obtained from (h-h) correlations, $W_{sat}$ as a function of $L$ ,
 and structure factor. 
The (h-h) correlation is 
\begin{eqnarray}
G(x,t)&=&\frac{1}{N}\sum_{x'}(h(x+x',t)-h(x',t))^{2}\nonumber \\
&=&x^{2\alpha}f\left(\frac{x}{\xi(t)}\right)
\end{eqnarray}
                                                                                                    
where, correlation length $\xi(t)\sim t^{1/z}$. In the limit $x\rightarrow
0$, $f\rightarrow 1$.

the width over a substrate of length $L$ as,
\begin{eqnarray}
w_{2}(L,t)&=&\frac{1}{N}\sum_{x}(h(x,t)-\bar h(t))^{2}\nonumber\\
&=&L^{2\alpha}g\left(\frac{L}{\xi(t)}\right),
\end{eqnarray}
Here $\bar h(t)$ is the average height at time $t$. 
It can be shown that \cite{bar} for times $t>>L$, $g\left(\frac{L}{\xi(t)}\right)
\rightarrow const.$ , thus $w_{sat} \propto L^{2\alpha}$. For $t<<L$ $w_{2}(t) \propto t^{2\beta}$.

The structure factor $S({\bf k},t)$ is measured as 
\begin{equation}
S({\bf k},t)=<h({\bf k},t)h(-{\bf k},t)> \nonumber\\
\end{equation}
where $h({\bf k},t)=(1/N)\sum_{\bf x} (h({\bf x},t)-\bar h({\bf x},t))e^{i{\bf k}.{\bf x}}$.

Figures (3), (4) and (5) summarize the results obtained for the model in (1+1) dimensions. 
 Fig. 3 shows the log-log plot of 
$G(x,t)$ Vs. $x$. The straight line is fitted between $x=$8 and 500. The slope gives 
$\alpha=0.504\pm 0.002$. Fig. 4 shows plot of  $W_{sat}$ Vs. $L$ 
for (1+1) dimensions. The least square fit
to the points gives $\alpha=0.502\pm0.005$. 

Fig. 5 shows log-log plot of $<h(k,t)h(- k,t)>$ Vs. $k$. The straight line fit is between $k$=0.03 
to 1.25. The slope near $k=\pi$ tends to zero \cite{sig}. The slope is 2.003$\pm 0.021$. 
This gives \cite{bar} $\alpha=0.500 \pm 0.021$.  Earlier, in reference \cite{mell} for 
the etching model, a slope 1.92 $\pm 0.02$ was obtained in the range of $k$=0.05 to 0.1, 
resulting in $\alpha=$0.46. This was considered to be one of the best value obtained 
for the existing lattice models by this method. {\it Clearly the present model 
provides a better value.}  
In the same reference, $\alpha=0.496$ is obtained from 
the $W_{sat}$. The apparent difference in the two $\alpha$ values indicates the 
presence of finite size effect for the etching model. In the present model, the slope is unaffected 
at smaller $k$ values. We have tested it up to $k=0.005$.

 Above results show that the proposed model has minimal finite size effects. It further 
confirms the method introduced in reference \cite{sve1} for determination of finite size 
effects in a lattice model. We have applied this method in (2+1) dimensions and in 
(3+1) dimensions to choose the model parameters corresponding to the minimal finite  
size effects. 

 In Fig. 6, we display the results of  $W_{sat}$ Vs. $L$ in (2+1) dimensions. The 
filled squares are the values calculated from the simulation results. The corresponding 
fit gives $\alpha=0.357\pm 0.005$. Fig. 7 shows the log-log plot of $G(x,t)$ Vs. $x$. 
From its slope,  $\alpha=0.355\pm 0.001$. The line is fitted between $x=$ 2 to 50. Thus 
both the methods give results matching within statistical margin. 

 In Fig. 6, we have also plotted the results of  $W_{sat}$ Vs. $L$ for the model in 
(3+1) dimensions. It gives  $\alpha=0.289\pm 0.005$. The $\alpha$ values obtained from 
these models in (2+1) and (3+1) dimensions are very close to those 
predicted in reference \cite{sve2}. In this 
reference $\alpha=0.35702$ for (2+1) dimensions and 0.28125 for (3+1) dimensions.

We have also measured the $\beta$ values for these models from the log-log plots of 
$w_{2}(t)$ Vs. $t$. We obtain $\beta= 0.332\pm 0.001$
,$0.221\pm 0.002$ and $0.168 \pm 0.003$ for (1+1), (2+1) and (3+1) dimensions respectively
. These values are consistent with the universal relation 
$\alpha +z=2$ for the KPZ equation \cite{bar}. 

 In conclusion, we have developed a lattice model belonging to the KPZ universality 
with minimum finite size effects. That the finite size effects are minimum is evident 
from the results obtained. The $\alpha$ value using $W_{sat}$ Vs. $L$ ,  $G(x,t)$ Vs. $x$
, and from $S(k,t)$ Vs. $k$ plots are equal within statistical margin in (1+1) dimensions. 
The simulation values are very close to the exact  $\alpha$, 0.5.  
Measurement involving local height fluctuations is successfully used in determining the finite size effects in lattice models . The models with minimum finite size effects 
are expected to lead to the better accuracy in determining the exact exponents for 
KPZ growth in higher dimensions. In (2+1) dimensions, we have obtained values close to 
$\alpha=0.36$ while in (3+1) dimensions it is around 0.29. Both these values are close to 
the earlier prediction in reference \cite{sve2}.

{}    
\pagebreak  

\begin{figure}
%\epsfxsize=\hsize \epsfysize = 3.0 in
%\centerline{\epsfbox{fgmdl1.eps}}
\label{Fig. 1}
\caption{Plot of distribution of  $\Delta h(t)_{local}$
for the (1+1) dimensional model discribed in the text on semi log scale. The distribution
 is for $t=5000$ MLs, with $\sigma=1.7$. The distribution is obtained by 
collecting the data over 3000 runs.  
}
\end{figure}

\begin{figure}
%\epsfxsize=\hsize \epsfysize = 2.0 in
%\centerline{\epsfbox{fgmdl1.eps}}
\label{Fig. 2}
\caption{Plot of $P_{0}$ in \% as a function of parameter $\sigma$ for the 
model in (1+1) dimensions. 
}
\end{figure}

\begin{figure}
%\epsfxsize=\hsize \epsfysize = 2.0 in
%\centerline{\epsfbox{fgcrit2.eps}}
\label{Fig. 3} 
\caption{Plot of $G(x,t)$ Vs. $x$ on log-log scale. The substrate size L=80000, 
and the number of monolayers grown are 5x10$^{5}$.  
}
\end{figure}

\begin{figure}
%\epsfxsize=\hsize \epsfysize = 3.0 in
%\centerline{\epsfbox{fgmdl3.eps}}
\label{Fig. 4} 
\caption{Plot of $W_{sat}$ Vs. $L$ on log-log scale for the model in (1+1) 
dimensions. 
}
\end{figure}

\begin{figure}
%\epsfxsize=\hsize \epsfysize = 3.0 in
%\centerline{\epsfbox{fgmdl4.eps}}
\label{Fig. 5} 
\caption{Plot of $S(k,t)$ Vs. $k$ on log-log scale for the model in (1+1) 
dimensions. The points are averaged over the substrate lengths of $L$=800, 
900,1000,1100,1200,1300,1350. $k$ varies from $\pi/100$ to $\pi$.  
}
\end{figure}

\begin{figure}
%\epsfxsize=\hsize \epsfysize = 3.0 in
%\centerline{\epsfbox{fgmdl5.eps}}
\label{Fig. 6} 
\caption{Plot of $W_{sat}$ Vs. $L$ on log-log scale for the model in (2+1) 
dimensions (filled squares) and for the model in (3+1) dimensions (filled circles). 
}
\end{figure}

\begin{figure}
%\epsfxsize=\hsize \epsfysize = 2.0 in
%\centerline{\epsfbox{fgmdl6.eps}}
\label{Fig. 7} 
\caption{Plot of $G(x,t)$ Vs. $x$ on log-log scale for the model in (2+1) dimensions.
 The substrate size is $L=800$, and the number of monolayers grown are 20000.  
}
\end{figure}

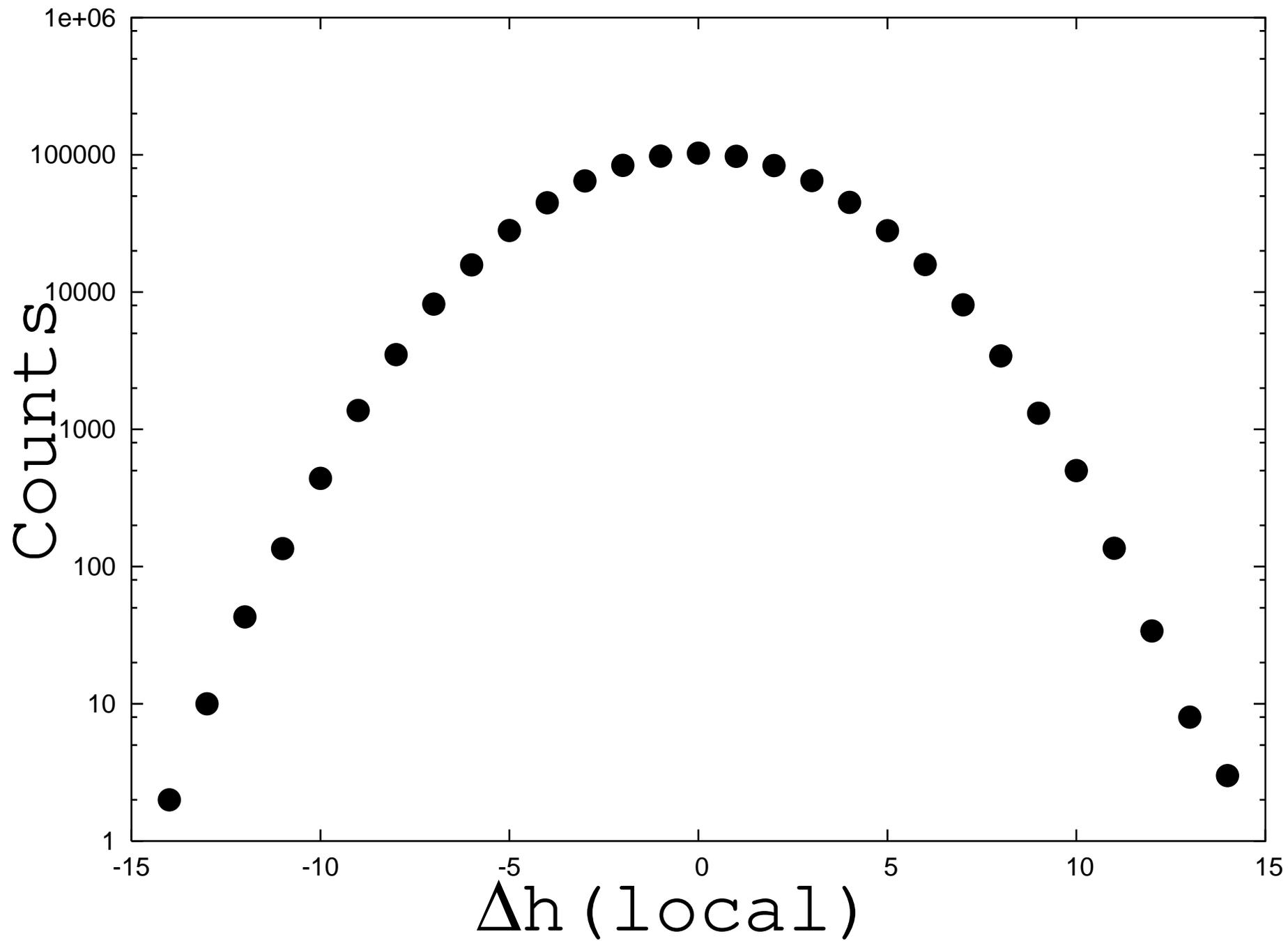
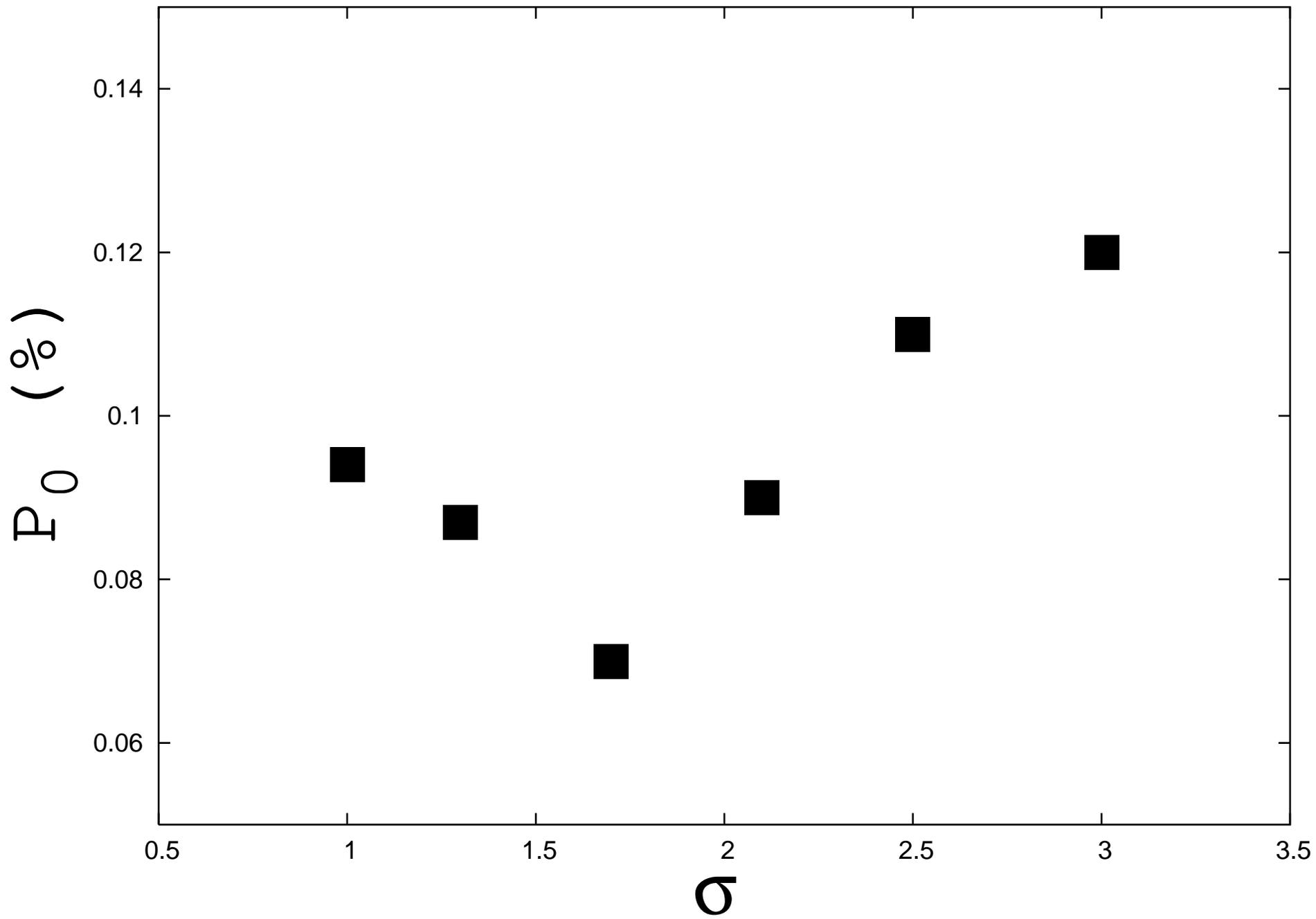
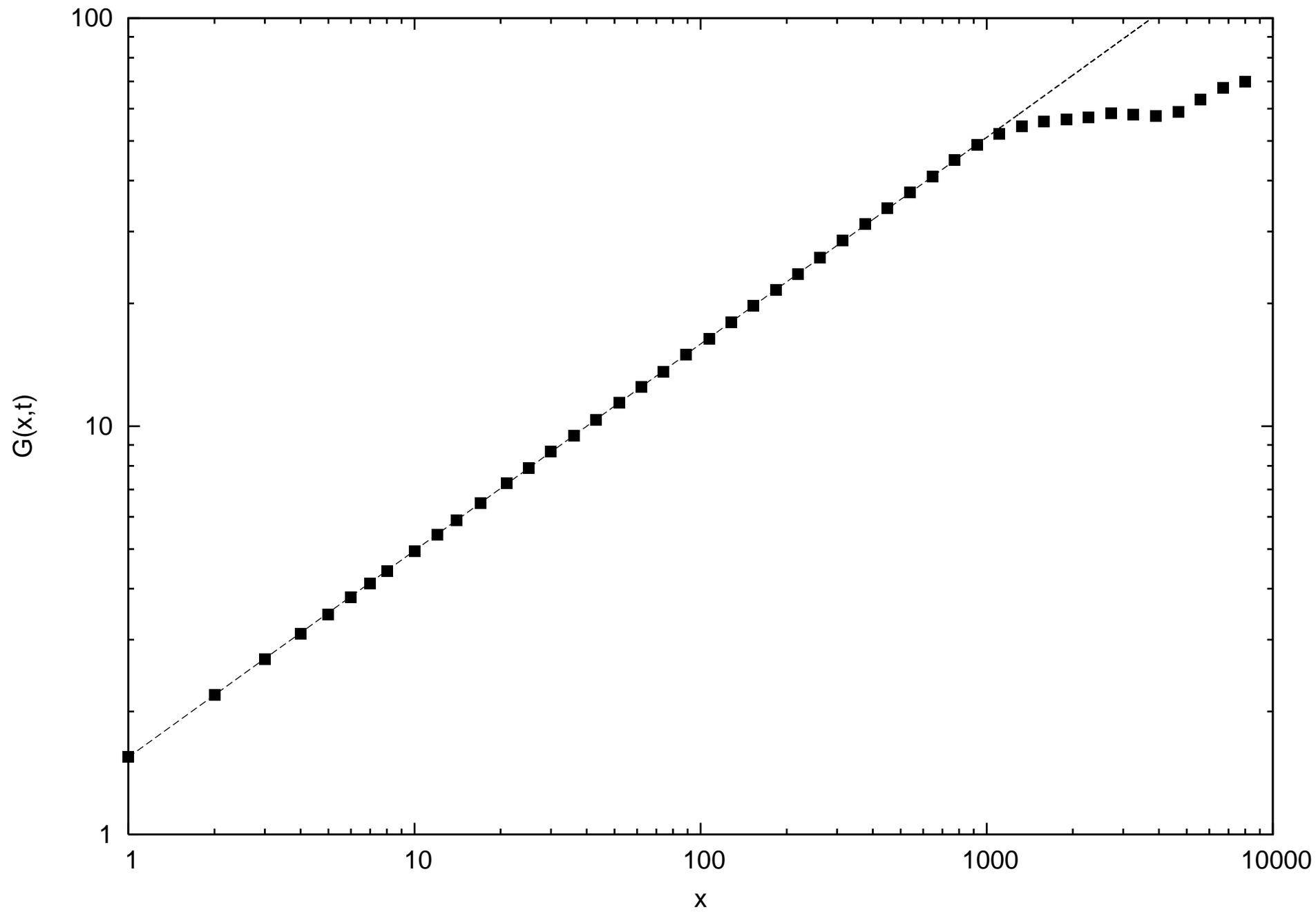
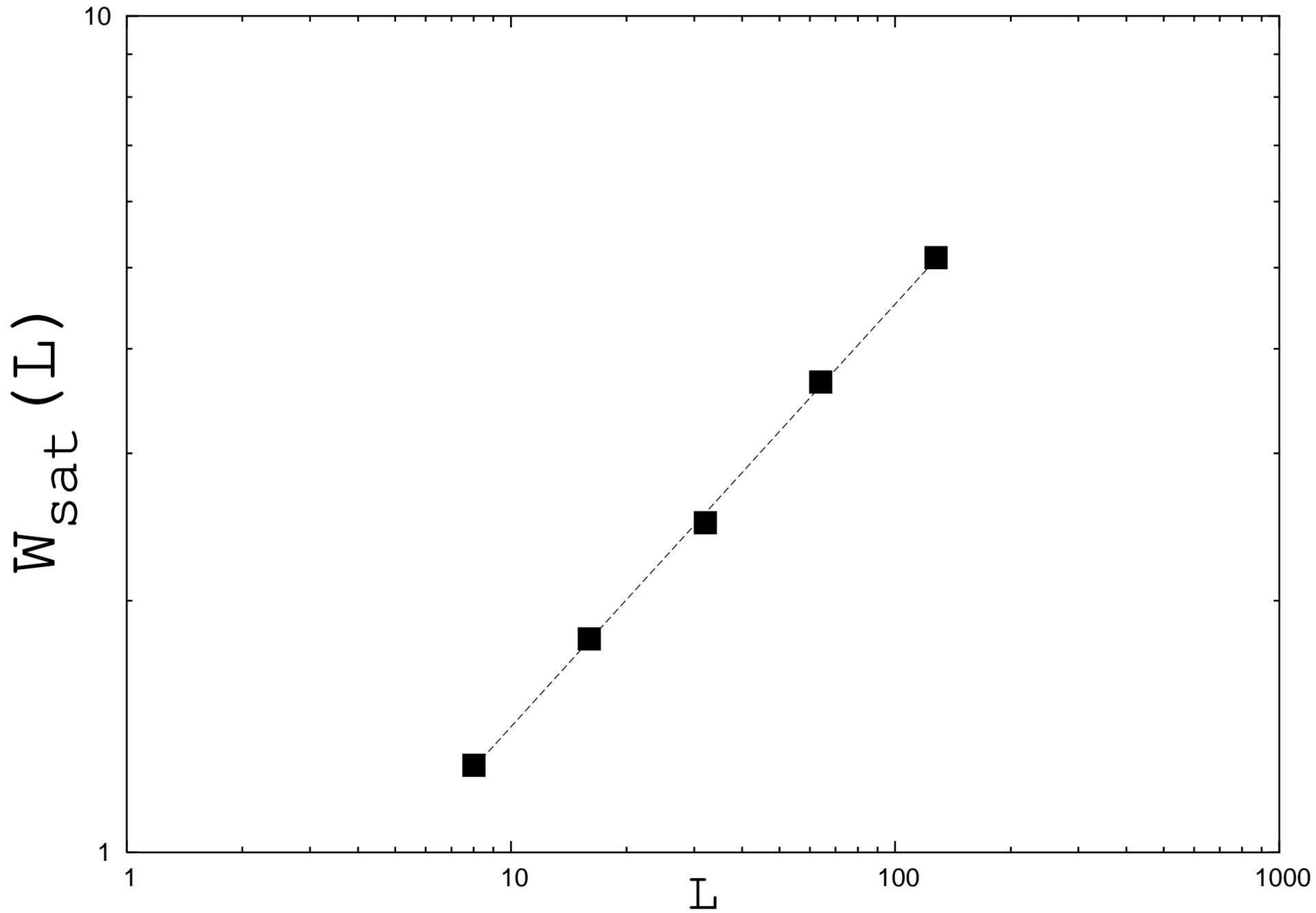
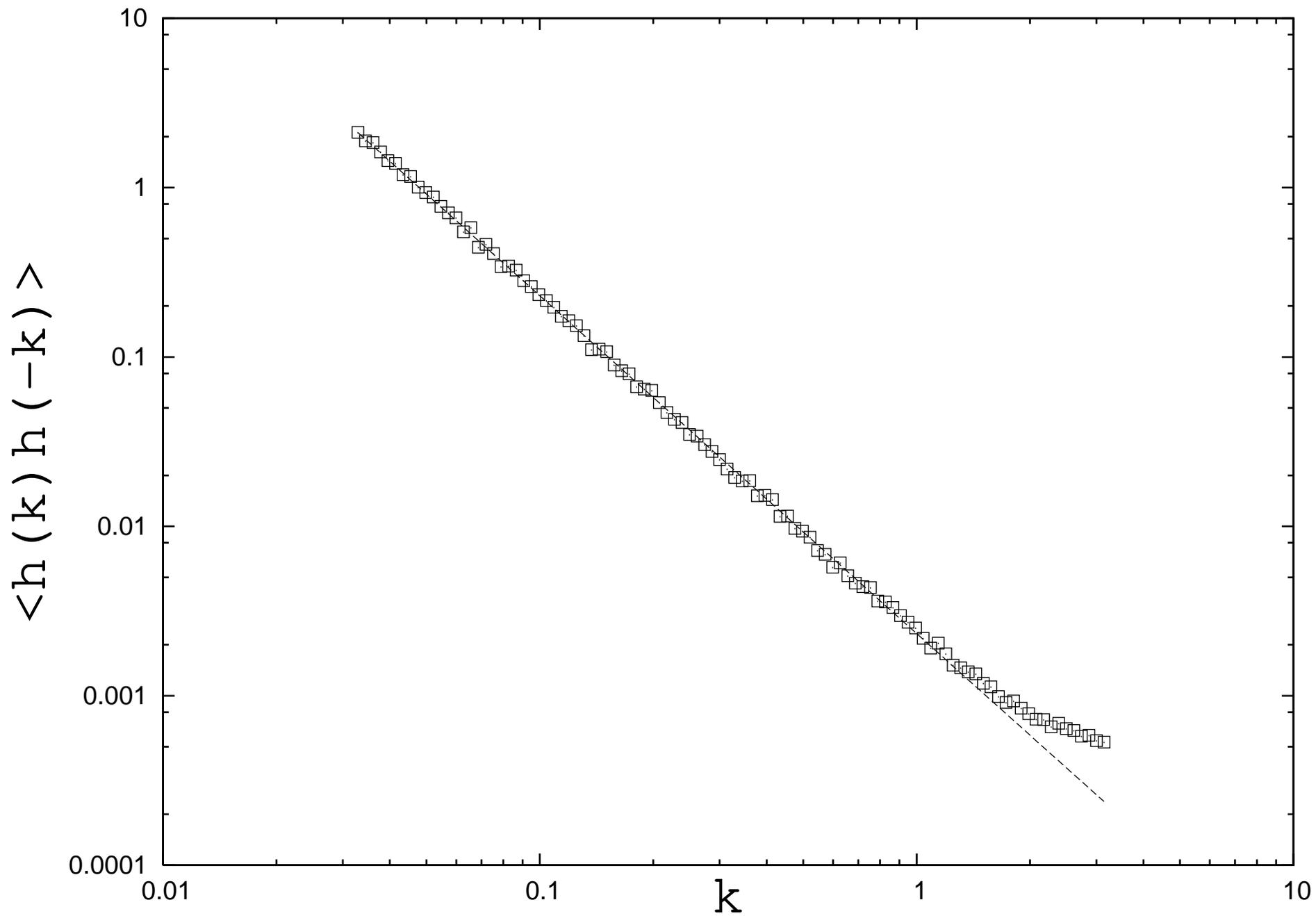
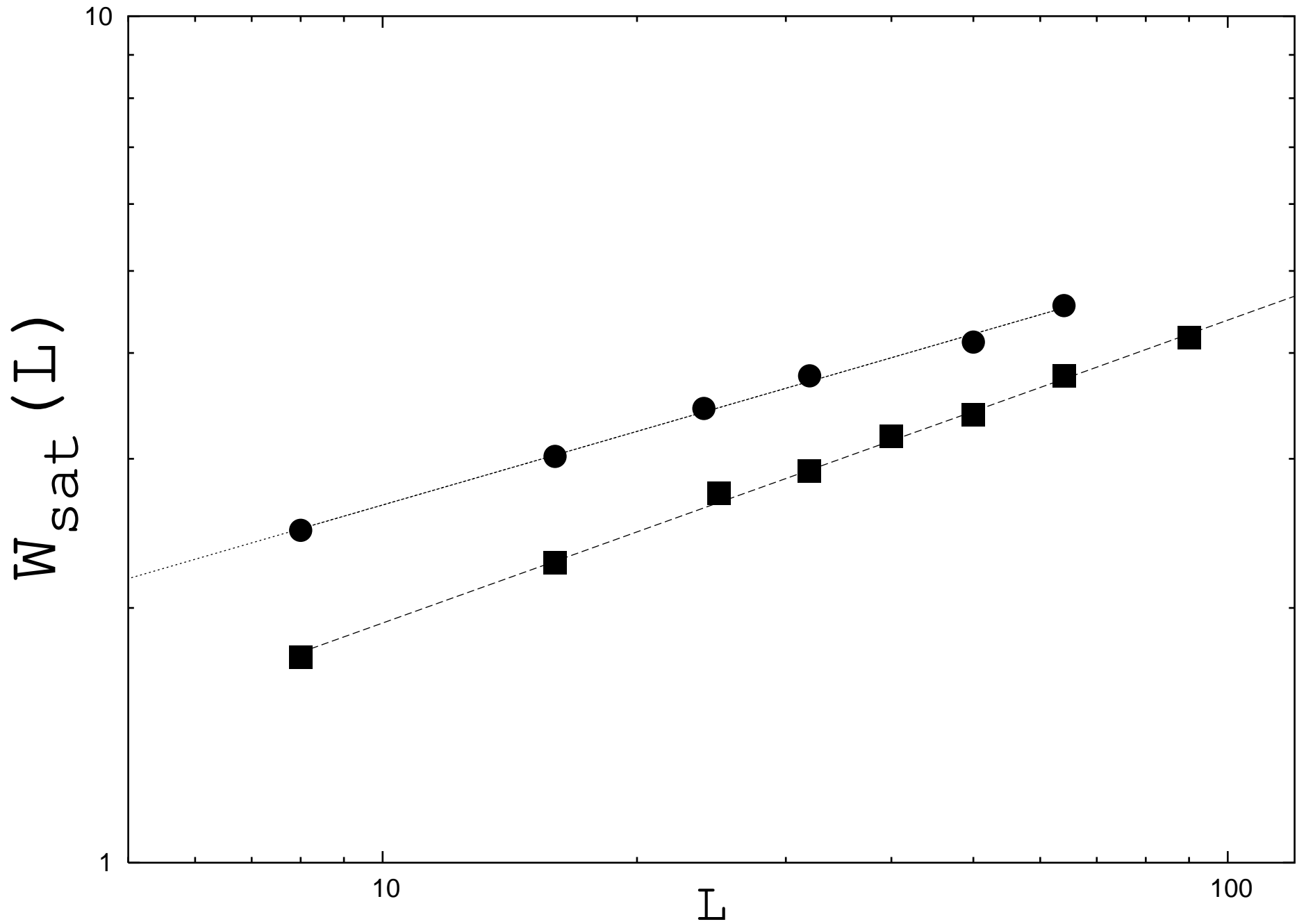
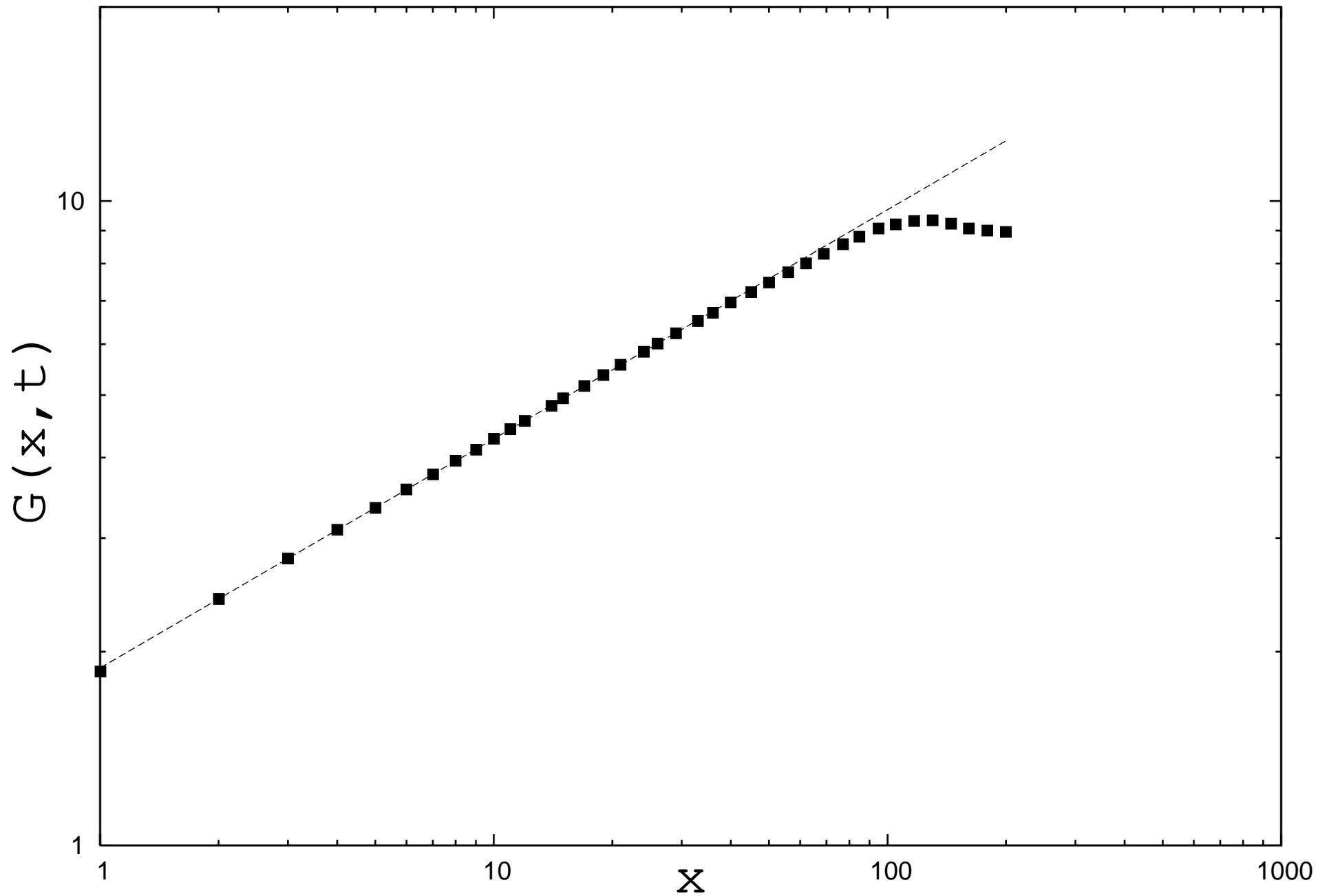
\end{document}